# Ptychographic Nanoscale Imaging of the Magnetoelectric Coupling in Freestanding BiFeO$_3$

*Tim A. Butcher,\* Nicholas W. Phillips, Chun-Chien Chiu, Chia-Chun Wei, Sheng-Zhu Ho, Yi-Chun Chen, Erik Fröjdh, Filippo Baruffaldi, Maria Carulla, Jiaguo Zhang, Anna Bergamaschi, Carlos A. F. Vaz, Armin Kleibert, Simone Finizio, Jan-Chi Yang, Shih-Wen Huang, and Jörg Raabe*

Understanding the magnetic and ferroelectric ordering of magnetoelectric multiferroic materials at the nanoscale necessitates a versatile imaging method with high spatial resolution. Here, soft X-ray ptychography is employed to simultaneously image the ferroelectric and antiferromagnetic domains in an 80 nm thin freestanding film of the room-temperature multiferroic BiFeO$_3$ (BFO). The antiferromagnetic spin cycloid of period 64 nm is resolved by reconstructing the corresponding resonant elastic X-ray scattering in real space and visualized together with mosaic-like ferroelectric domains in a linear dichroic contrast image at the Fe L$_3$ edge. The measurements reveal a near perfect coupling between the antiferromagnetic and ferroelectric ordering by which the propagation direction of the spin cycloid is locked orthogonally to the ferroelectric polarization. In addition, the study evinces both a preference for in-plane propagation of the spin cycloid and changes of the ferroelectric polarization by 71° between multiferroic domains in the epitaxial strain-free, freestanding BFO film. The results provide a direct visualization of the strong magnetoelectric coupling in BFO and of its fine multiferroic domain structure, emphasizing the potential of ptychographic imaging for the study of multiferroics and non-collinear magnetic materials with soft X-rays.

## 1. Introduction

Among the discovered single-phase multiferroics,[1] BiFeO$_3$ (BFO) stands out because its multiferroicity persists at high temperatures with a ferroelectric Curie temperature of $T_C = 1110$ K and an antiferromagnetic Néel temperature of $T_N = 640$ K. The magnetoelectric coupling in BFO is promising for technological applications, since controlling magnetism with an electrical field instead of spin-polarized electrical currents would boost efficiency.[2–4] Additionally, antiferromagnetic order is enticing because magnetic perturbations and stray fields are avoided. Antiferromagnets also promise the operation of devices at terahertz frequencies, surpassing the gigahertz range to which ferromagnets are restricted.[5,6]

The ferroelectricity of BFO is of lone-pair type with a large polarization ($P \approx 90\,\mu$C cm$^{-2}$) along the pseudocubic ⟨111⟩ directions (see unit cell in **Figure 1a**). Originally, it was thought that the magnetic Fe$^{3+}$ (d$^5$) ions were ordered antiferromagnetically in a G-type configuration.[7] However, a neutron diffraction study by Sosnowska et al. in 1982[8] showed that the magnetic moments are in fact non-collinear and twist in form of a spin cycloid with a period of approximately 64 nm (see Figure 1b). The cycloidal ordering is a manifestation of the Dzyaloshinskii–Moriya interaction (DMI) and is incommensurate with the crystal lattice.[9] The cycloid can propagate in three directions that are defined by orthogonality to the ferroelectric polarization.

Ferroic materials minimise energy by forming domains that are critical for the reaction of the system to external fields. This is particularly pertinent in multiferroic BFO, where ferroelectric polarization and antiferromagnetic order influence each other by magnetoelectric coupling. The ferroelectric polarization in neighboring BFO domains can change by 71°, 109°, or 180° for the eight equivalent ⟨111⟩ directions.[10]

While the imaging of ferroelectric domains is routinely carried out with piezoresponse force microscopy (PFM), antiferromagnetic domains are more challenging to access experimentally.[11] The presence of non-collinear periodic antiferromagnetic

T. A. Butcher, N. W. Phillips, E. Fröjdh, F. Baruffaldi, M. Carulla, J. Zhang, A. Bergamaschi, C. A. F. Vaz, A. Kleibert, S. Finizio, S.-W. Huang, J. Raabe
Paul Scherrer Institut
Villigen PSI 5232, Switzerland
E-mail: tim.butcher@psi.ch

C.-C. Chiu, C.-C. Wei, S.-Z. Ho, Y.-C. Chen, J.-C. Yang
Department of Physics
National Cheng Kung University
Tainan 70101, Taiwan

J.-C. Yang
Center for Quantum Frontiers of Research & Technology (QFort)
National Cheng Kung University
Tainan 70101, Taiwan

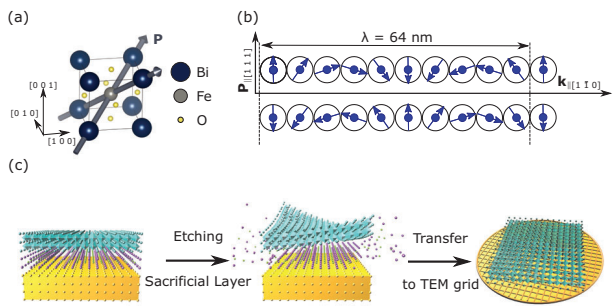

**Figure 1.** Freestanding multiferroic BFO films. a) Pseudocubic unit cell of BFO with the ferroelectric polarization (**P**) along [111] and [1$\bar{1}$1]. There is an angle of 71° between these directions. b) The magnetic moments of the $Fe^{3+}$ ions are coupled antiferromagnetically and form a spin cycloid that propagates orthogonally to **P**. c) The BFO film (cyan) was grown along the [001] axis on a (001) $SrTiO_3$ substrate (yellow) (see Experimental Section). The freestanding 80 nm thin BFO film was obtained by etching the (001) $Sr_3Al_2O_6$ (SAO) sacrificial layer (purple), after which it was transferred to a TEM grid.

ordering further complicates the visualization of the magnetic order in view of the requirement of high spatial resolution. Previously, the spin cycloid in BFO has only been imaged by scanning nitrogen-vacancy (NV) magnetometry.[12–17] This surface sensitive technique was used to map the weak stray field that results from uncompensated spins due to a second DMI interaction.[18] Recently, ferroelectric domains have been observed by measuring the Stark shift of the NV spin with a resolution of ≈100 nm.[19] Yet, a high resolution imaging method sensitive to both ferroelectric and magnetic order in a single measurement is hitherto absent, despite being key for the study of multiferroics. The present study applies the synchrotron technique of soft X-ray ptychography[20–23] to accomplish this in a freestanding thin film of BFO.

Synchrotron X-rays are a powerful tool for imaging multiferroics, since in-plane antiferromagnetic and ferroelectric ordering can be visualized by toggling the linear polarization to realize contrast as X-ray linear dichroism (XLD). For the 3d magnetic transition metals, the L-absorption edges can exhibit strong XLD because of antiferromagnetism or ferroelectricity. The surface sensitive technique of photo-emission electron microscopy (X-PEEM) has featured prominently in the study of BFO since the 2000s.[10,24] A seminal study was able to show the coupling between ferroelectric and antiferromagnetic domains by heating above $T_N$.[10] Nonetheless, X-PEEM studies have been impeded by spatial resolutions in the order of 50 nm that are insufficient to resolve modulations of the spins, such as the spin cycloid of BFO. The method of scanning transmission X-ray microscopy (STXM), in which the sample is scanned through the X-rays in the focal point of a Fresnel zone plate while the X-ray transmission is recorded, improves upon the resolution of X-PEEM. However, lithographically fabricated Fresnel zone plates limit the resolution for STXM, which routinely approaches 25 nm and is insufficient to image the spin cycloid of BFO.

Ptychographic imaging approaches[25] overcome the limitations in spatial resolution faced by other synchrotron imaging methods. Whilst geometrically similar to STXM instrumentation, ptychography makes use of a pixelated area detector, rather than a point detector. Data collection consists of measuring coherent diffraction patterns from overlapping regions on the sample. The beam spot size is controlled by a Fresnel zone plate in what is known as defocused probe ptychography[26] (see Experimental Section). Both the amplitude and phase information of the sample and illumination are recovered by an iterative phase retrieval algorithm.[27] Unlike hard X-ray ptychography,[28,29] soft X-rays provide strong dichroic contrasts in ferroic 3d transition metal compounds[21,22] such as BFO, since the relevant atomic transitions lie within the range from 700 to 900 eV. Furthermore, as a coherent diffractive imaging technique, ptychography ties in with X-ray diffraction studies, which have contributed crucially to the study of non-collinear magnetic ordering in multiferroics.[30]

## 2. Results

The analyzed BFO sample was an 80 nm thin freestanding film grown along the [001] direction (see Experimental Section). This was transferred from a $SrTiO_3$ substrate to a transmission electron microscopy (TEM) grid with a lacey carbon supported film, which was achieved by means of a water-soluble sacrificial layer (see Figure 1c). A TEM image indicating the high quality of the crystalline film is shown in **Figure** 2b. While epitaxially grown BFO on substrates such as $SrTiO_3$ or rare earth scandates have been widely studied since the 2000s, freestanding oxide films are a comparatively new system that can be transferred to any support.[31–35] In the case of BFO, it has been shown that the absence of a substrate can lower the switching voltage and improve the switching speed of the ferroelectric polarization.[33] Freestanding films are also well suited for soft X-ray ptychography, considering that the absorption at resonance limits the sample thickness to a few hundred nanometers.

The XLD induced by the ferroelectric polarization follows a (1 − $\cos^2\theta$) variation with $\theta$ as the angle between the electric field of the X-rays and the projection of the ferroelectric polarization.[36] The electric field of the X-rays used here lies in the (001) plane of the BFO freestanding film. Only domains in which the ferroelectric polarization changes by 71° or 109° can be visualized by XLD in BFO, while changes by 180° remain indistinguishable. The orientation of the BFO sample on the ptychography sample holder was determined from the TEM image and is sketched in Figure 2c, which shows one in-plane projection of the ferroelectric polarization (**P**), cycloid propagation vector (**k**), horizontal and vertical polarization ($L_H$ and $L_V$) directions of the X-rays. This arrangement of the sample resulted in $\theta = 70°$ or $\theta = 20°$ with respect to $L_H$ depending on the orientation of **P** in the corresponding domain.

When the BFO is exposed to linearly polarized X-rays with energies tuned to the Fe $L_3$ edge (see Supporting Information for XLD spectrum from STXM measurements), the periodic spin cycloid reveals itself as two diffraction peaks, which can be seen in Figure 2d for a focused beam. They appear as a convolution of the cycloidal diffraction peak with the divergent illumination from the Fresnel zone plate. The central position of the diffraction peak corresponds to half the cycloid period of 64 nm. This is because the magnetic resonant X-ray scattering is proportional to $m^2$,[13] $m$ being the projection of the magnetic moments along the electric field direction. From the measurement of a single projection it is only possible to determine the magnitude of $m$, i.e., the sign remains unknown and this is reflected in the reconstructions of the ptychographic measurement in which the spin cycloids emerge with half their period. The reconstructed phase image with $L_H$ polarization is displayed in Figure 2e. Aside from the lacey carbon support film of the TEM grid, numerous domains containing spin cycloids as stripe-like features in the [110] or [1$\bar{1}$0] directions are discernible.

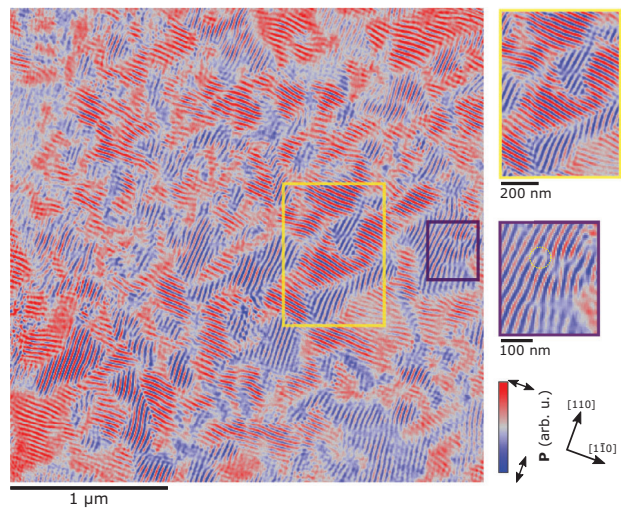

Figure 3. Linear dichroic phase contrast image with multiferroic domains. The mosaic-like ferroelectric domains are in red and blue color. Additionally, the XLD phase image shows the stripes due to the periodic cycloidal modulation of the magnetic order. The orientation of the ferroelectric domains dictates the propagation direction of the cycloid. The magnified inset of the image framed in yellow shows several multiferroic domains and the change of the cycloid direction at the domain walls. The lower inset shows a domain with magnetic defects in the form of cycloidal edge dislocations in which the cycloid forks into two parts. One is encircled in yellow.

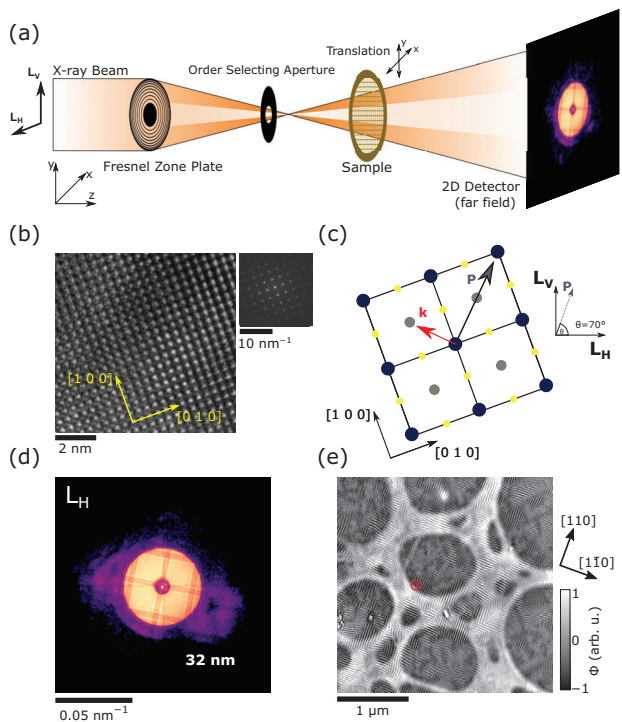

**Figure 2.** Soft X-ray ptychography of BFO. a) Sketch of the soft X-ray ptychography setup: A monochromatic X-ray beam is focused before the sample, which is moved through the beam using a piezoelectric stage. Diffraction patterns are recorded from multiple, overlapping regions on the sample. b) TEM image of the BFO freestanding film and respective electron diffraction pattern. c) Relative alignment of the BFO unit cell with experimental geometry showing the ferroelectric polarization (**P**), cycloid propagation vector (**k**), $L_H$ and $L_V$ linear X-ray polarizations. The principal crystallographic axes of the sample were rotated in-plane by 20° with respect to the direction of oscillation of the X-ray electric field. d) Diffraction pattern at the Fe $L_3$ edge with a focused beam ($L_H$). The cycloid gives rise to two diffraction peaks on each side of the central Fresnel zone plate cone. e) Reconstructed phase image from a ptychography scan. Many domains with cycloids are visible as stripey patterns on the lacey carbon support film. The position at which the diffraction pattern in (d) was recorded is encircled in red.

Domains with vertically propagating cycloids are brighter than the horizontal case in the phase image captured with $L_H$ polarized X-rays. This is due to differences in the absorption of linearly polarized X-rays caused by asymmetries in the charge distribution stemming from the ferroelectric domains. The direct imaging of the domain configuration without the topography of the sample is achieved by subtracting the phase images taken at $L_H$ and $L_V$ polarizations. The resulting linear dichroic phase image is shown in **Figure 3**. Both the ferroelectric and antiferromagnetic order of BFO are distinctly perceivable in this image. On the one hand, the different projections of the ferroelectric polarization appear as red and blue domains owing to the XLD. On the other hand, the image shows the antiferromagnetic spin cycloids, which manifest themselves as small periodic stripey ripples superimposed on the ferroelectric domains. For linearly polarized light, the aforementioned $(1 - \cos^2\theta)$ dependence of the ferroelectric signal allows for unambiguous separation from the antiferromagnetic cycloidal modulation whose diffraction peak exists regardless of the angle between **k** and the electric field of the X-rays.

The cycloids in the freestanding film propagate exclusively within the plane of the film, which is akin to the behavior in thin films on substrates with low epitaxial strain.[15] What is more, there is a perfect correspondence between the ferroelectric domains and the cycloid propagation direction, demonstrating the strong magnetoelectric coupling in BFO.

The ferroelectric domains are mosaic-like and range in size from below 100 nm to above 600 nm in length. These are predominately separated by 71° domain walls (see ferroelectric polarization directions in Figure 1a), which was concluded from PFM measurements (see Supporting Information). The out-of-plane components of the ferroelectric polarization point only in one direction in this configuration, which means that the associated stray field is substantial.

The spatial resolution within a multiferroic domain is as high as 4.9 nm, based on Fourier ring correlation (see Supporting Information). This allows the direct observation of the alteration of the cycloid direction at the multiferroic domain walls (see upper magnified area of Figure 3). It can be appreciated that the spin cycloids are not straight in every domain and instead curve slightly. Part of the reason for this lies in the magnetoelectric coupling, which forces the spins in the cycloids to align when they change propagation direction abruptly by 90° amid transition through a domain wall separating the irregularly sized domains.

Imperfections in the cycloidal order also occur as cycloidal edge dislocations in which two branches of the cycloid merge into one. These are often present in the vicinity of the domain walls, but also nucleate further in the center of the domains. This can be seen in the bottom zoomed-in area of Figure 3. Such defects have been identified in helical systems[37,38] and have also been uncovered in the surface layer of a bulk BFO crystal with NV magnetometry.[16] It is to be expected that changes of the ferroelectric domain structure of BFO by external fields will modify the defects. Apart from highlighting the disturbances in the cycloidal order and directly relating it to the multiferroic domain structure, the high resolution images also evince that the antiferromagnetic order and magnetoelectric coupling is maintained even in small islands of multiferroic domains with dimensions approaching a single period of the cycloid. On a final note, the cycloidal antiferromagnetic order can be isolated from the ferroelectric textures by exploiting circular dichroism in the X-ray scattering,[13,14] which originates in the cycloidal rotation direction in each domain (see Supporting Information).

## 3. Conclusion

Soft X-ray ptychography is the first technique to image the spin cycloid and the ferroelectric domains in BFO concurrently. The results provide direct evidence of strong magnetoelectric coupling, with the technique being well suited to probe other multiferroic materials at the nanoscale. Soft X-ray ptychography is element specific, bulk sensitive and adaptable to a range of sample environments including temperature, magnetic field and strain. The ability to efficiently scan different areas of the sample is of high importance for high-throughput studies that are unrestricted to specific regions. Besides this, the sensitivity to the bulk properties in the transmission geometry provides unique information that cannot be obtained by surface sensitive microscopic techniques alone.[16] It also synergizes with the elemental selectivity with which buried layers can be accessed. Thanks to the similarity of the requirements for samples in soft X-ray ptychography and TEM, correlative studies in which multiferroic domain walls selected by ptychography can subsequently be imaged on an atomistic level by TEM are feasible. The ptychographic imaging process with soft X-rays may be extended to three dimensions by rotating samples in a laminographic geometry.[39] This would remove issues due to the equivalence of the projections for 71° and 109° domain walls when imaging BFO with the X-ray beam parallel to the [001] direction. Up to now, the 3D imaging of ferroelectric domain walls has only been accomplished with the destructive method of focused ion beam-scanning electron microscopy (FIB-SEM) tomography.[40] Finally, it is envisaged that soft X-ray ptychography will enable high resolution time resolved studies of the effect of external stimuli on ferroics in a pump-probe setup.[29,41]

## 4. Experimental Section

*Thin Film Growth and Lift-Off Procedure*: The freestanding BFO samples were obtained by etching the $Sr_3Al_2O_6$ (SAO) sacrificial layer of epitaxial BFO/SAO heterostructures grown on (001) $SrTiO_3$ (STO) substrates. The epitaxial BFO/SAO bilayers were fabricated using pulsed laser deposition (Mobile Combi-Laser MBE MC-LMBE, Pascal Co., Ltd.). All the growth processes utilized a KrF excimer laser with a power of 200 mJ and a frequency of 10 Hz. The fabrication process commenced with the deposition of a 25 nm SAO layer on the STO substrate at 715° C and an oxygen pressure of 50 mTorr. Subsequently, an 80 nm BFO layer was deposited on top of the SAO layer at 700° C and an oxygen pressure of 100 mTorr. The sample was then annealed at 400° C under near 1 atm oxygen pressure for 30 min to minimize oxygen vacancies. Gradual cooling to room temperature was performed while maintaining the same oxygen pressure. Finally, the sacrificial SAO layer was dissolved in deionized water, enabling the transfer of the freestanding BFO film from the solution to a TEM grid for further analysis (see Figure 1c). The orientation of the resulting BFO freestanding film was determined by localized electron diffraction in TEM (see Figure 2b).

*Piezoresponse Force Microscopy*: The ferroelectric domains of a sample from the same batch of freestanding films placed on Si were measured by PFM (see Supporting Information). The topography and PFM images were acquired using a modified scanning probe microscope system (Multimode 8, Bruker) with a Nanoscope Controller V. PFM measurements were carried out under the dual-frequency-resonance-tracking (DFRT) mode to reduce the cross-talk effect from the morphology, employing commercial Pt/Ir-coated tips with a spring constant of 2.8 N m$^{-1}$ (NANOSENSORS PPP-EFM60). During the DFRT-PFM measurement, the tip was driven with a primary AC voltage of an amplitude of approximately 1.5 V and tracked by a lock-in amplifier (HF2LI, Zurich Instruments) at the contact-resonance frequency of about 650 kHz.

*Soft X-Ray Ptychography*: Ptychographic imaging with soft X-rays was performed at the Surface/Interface Microscopy (SIM) beamline at the Swiss Light Source. Linearly or circularly polarized X-rays were provided by an elliptical Apple II UE56 undulator. The sample was placed 96 mm in front of a 512×512 pixel single photon counting soft X-ray detector with a pixel size of 75 $\mu$m. An exposure time of 200 ms was used.

The soft X-ray detector comprised four EIGER readout chips[42] bump bonded to a single Low Gain Avalanche Diode (LGAD) sensor optimized for soft X-ray detection.[43] The sensor was developed in-house in collaboration with Fondazione Bruno Kessler (Trento, Italy) and was based on the inverse LGAD technology,[44] which enables single photon counting below 1 keV that was typically hindered by the electronic noise of the readout electronics.[45] The new detector provided the same low noise, large dynamic range and fast frame rate performance that hard X-ray photon counting detectors had relied on for over a decade. The optimization of the entrance window[46] and of the gain layer provided a high quantum efficiency and single photon resolution down to approximately 500 eV.

A 20 nm outer zone width Ir zone doubled Fresnel zoneplate with 500 $\mu$m diameter was employed to focus the monochromatic X-ray beam. A 75 $\mu$m diameter order selecting aperture (OSA) close to the sample was employed to suppress X-rays from higher diffraction orders, while a 2 $\mu$m thick Au central stop with 100 $\mu$m diameter was used to block the intense undiffracted central beam. Ptychography scans consisted of moving the sample through the 900 nm FWHM beam on a Fermat spiral trajectory with steps of approximately 100 nm. Reliable positioning of the sample with 5 nm accuracy was guaranteed by a differential heterodyne laser interferometer coupled to the piezoelectric sample stage.

The object and illuminating wavefront were reconstructed with the Ptychoshelves software package[47] using 1200 iterations of difference-map[48] followed by 600 iterations of maximum-likelihood refinement.[49] Three probe modes were used in the reconstruction.[50]

## Supporting Information

Supporting Information is available from the Wiley Online Library or from the author.

## Acknowledgements

Soft X-ray ptychography measurements were performed at the Surface/Interface Microscopy (SIM-X11MA) beamline of the Swiss Light

Source (SLS), Paul Scherrer Institut, Villigen, Switzerland. T.A.B. acknowledged funding from the Swiss Nanoscience Institute (SNI). N.W.P. received funding from the European Union's Horizon 2020 research and innovation programme under the Marie Skłodowska-Curie grant agreement no. 884104. J.-C.Y. acknowledged the financial support from National Science and Technology Council (NSTC) in Taiwan, under grant no. NSTC 112-2112-M-006-020-MY3. The authors also thank MSSORPS Co., Ltd. for high-quality TEM sample preparation and preliminary examination. The research was also supported in part by Higher Education Sprout Project, Center for Quantum Frontiers of Research & Technology (QFort) at National Cheng Kung University, Taiwan. Scanning transmission X-ray microscopy measurements and X-ray characterization of the freestanding film were performed at the PolLux (X07DA) and MS (X04SA) beamlines of the Swiss Light Source, Paul Scherrer Institut, Villigen, Switzerland. The PolLux endstation was financed by the German Ministerium für Bildung und Forschung (BMBF) through contracts 05K16WED and 05K19WE2. The LGAD sensors were fabricated at Fondazione Bruno Kessler (Trento, Italy).

Open access funding provided by ETH-Bereich Forschungsanstalten.

## Author Contributions

T.A.B. and S.-W.H conceived the project. T.A.B. and N.W.P. performed ptychography measurements and analyzed the data. C.-C.C., C.-C.W., S.-Z.H., Y.-C.C., J.-C.Y., and S.-W.H. provided and characterized the freestanding film. T.A.B., N.W.P., C.A.F.V, A.K., S.F., J.-C.Y., S.-W.H., and J.R. discussed the data interpretation. J.R., S.F., T.A.B., C.A.F.V., and A.K. developed ptychography instrumentation. E.F., F.B., M.C., J.Z., and A.B. developed the detector. T.A.B. wrote the manuscript with input from all co-authors.

## Conflict of Interest

The authors declare no conflict of interest.

## Data Availability Statement

The data that support the findings of this study are available from the corresponding author upon reasonable request.

## Keywords

bismuth ferrite, freestanding films, multiferroic domains, soft X-ray ptychography, synchrotron techniques, X-ray microscopy